\documentclass[12pt,preprint]{aastex}

\newcommand{\rosat}{{\sl ROSAT\/}}

\newcommand{\sax}{{\sl Beppo-SAX\/}}
\newcommand{\xte}{{\sl RXTE\/}}

\newcommand{\xmm}{{\sl XMM-Newton\/}}

\newcommand{\nustar}{{\sl NuSTAR\/}}

\newcommand{\nh}{N$_{\rm H}$}

\shorttitle{Reflection in Intermediate Polars}
\shortauthors{Mukai et al.}

\begin{document}


\title{Unambiguous Detection of Reflection in Magnetic Cataclysmic Variables:
       Joint \nustar\ -- \xmm\ Observations of Three Intermediate Polars}


\author{K. Mukai\altaffilmark{1}}
\affil{CRESST and X-ray Astrophysics Laboratory, NASA/GSFC, Greenbelt,
       MD 20771}
\email{Koji.Mukai@nasa.gov}

\author{V. Rana}
\affil{Cahill Center for Astronomy and Astrophysics,
       California Institute of Technology, Pasadena, CA 91125}
\author{F. Bernardini}
\affil{New York University Abu Dhabi, P.O. Box 129188, Abu Dhabi, United Arab Emirates}
\author{D. de Martino}
\affil{INAF -- Osservatorio Astronomico di Capodimonte, Salita Moiariello 16,
       I-80131 Napoli, Italy}


\altaffiltext{1}{Also Department of Physics, University of Maryland,
       Baltimore County, 1000 Hilltop Circle, Baltimore, MD 21250.}


\begin{abstract}
In magnetic cataclysmic variables (CVs), X-ray emission regions are located
close to the white dwarf surface, which is expected to reflect a significant
fraction of intrinsic X-rays above 10 keV, producing a Compton reflection
hump. However, up to now, a secure detection of this effect in magnetic
CVs has largely proved elusive because of the limited sensitivity of
non-imaging X-ray detectors. Here we report our analysis of joint
\nustar/\xmm\ observations of three magnetic CVs, V709~Cas, NY~Lup, and
V1223~Sgr. The improved hard X-ray sensitivity of the imaging \nustar\ 
data has resulted in the first robust detection of Compton hump in all
three objects, with amplitudes of $\sim$1 or greater in NY~Lup, and likely
$<$1.0 in the other two. We also confirm earlier report of a strong spin
modulation above 10 keV in V709~Cas, and report the first detection of small
spin amplitudes in the others.  We interpret this as due to different height
of the X-ray emitting region among these objects. A height of $\sim$0.2
white dwarf radii provides a plausible explanation for the low reflection
amplitude of V709~Cas. Since emission regions above both poles are visible
at certain spin phases, this can also explain the strong hard X-ray spin
modulation. A shock height of $\sim$0.05 white dwarf radii can explain our
results on V1223~Sgr, while the shock height in NY~Lup appears negligible.
\end{abstract}





\section{Introduction}

In magnetic cataclysmic variables (CVs), a strongly magnetic ($\sim$1--200 MG)
white dwarf primary accretes matter from a Roche-lobe filling secondary.
Intermediate polars (IPs) are a subset of magnetic CVs in which the spin period
is significantly shorter than the orbital period. When the magnetically
controlled accretion flow hits the white dwarf surface, a strong standing
shock forms. The predominant cooling mechanism of shock-heated plasma in IPs
is the collisionally excited, thermal, multi-temperature X-ray emission
\citep{A1973}. Since the shock is formed near the stellar surface,
we expect that roughly half the intrinsic emission is directed towards
the primary and is reprocessed and/or reflected. The strongest
spectral signatures of reflection are the 10--30 keV Compton reflection
hump and the Fe K fluorescent line at 6.4 keV. While the latter are routinely
detected from magnetic CVs, they can also originate from intrinsic absorbers.
The importance of reflection has nevertheless been inferred through a
statistical analysis of the equivalent width of the 6.4 keV line vs.
absorbing column \citep{EI1999}.

A direct detection of the Compton reflection hump in magnetic CVs has
largely been elusive to date, due to a lack of X-ray spectra
of sufficient quality above 10 keV. Positive detections are reported
for AM~Her \citep{Retal1981,Betal1995,Metal2000} and for EF~Eri
\citep{Detal1995}. However, the studies on AM~Her used
single-temperature models as the basis; \cite{Detal1995}
also started with this model, and although they also explored reflection
of the expected multi-temperature emission, the data were not good enough
to distinguish between the two.
%When \cite{Cetal1998} applied their
%multi-temperature spectral model to the {\sl Ginga\/} spectra of magnetic
%CVs, they found that fits with and without reflection, as calculated by
%\cite{vTetal1996}, were roughly comparable.
\cite{Betal2000} found
residuals around the Fe K complex in their study of V1223~Sgr using
{\sl Ginga\/} data, and interpreted it as due to reflection. However,
they were unable to verify this through the detection of the
characteristic spectral curvature of reflection on the higher
energy side of the 10--30 keV hump.
% \cite{Setal2005} studied RXTE spectra without including reflection.

Previous hard X-ray data on these magnetic CVs generally had poor
signal-to-noise ratios and suffered from systematic uncertainties
of background models. With the advent of \nustar, the
first mission to extend the energy range for imaging X-ray observations
up to $\sim$79 keV, we can obtain much improved hard X-ray spectra. It
also enables a reliable study of spin modulation of X-rays above 10 keV.
In this Letter, we present the initial results from our joint observations
of 3 IPs using \nustar\ \citep{NUSTARREF} and \xmm\ \citep{XMMREF}, and
report the unambiguous detection of Compton reflection hump in all three
targets.

\section{Observations and Data Reduction}

We have selected V709~Cas, NY~Lup, and V1223~Sgr as our targets
for this study, because they have the highest observed 14--195 keV
{\sl Swift\/}/BAT fluxes according to \cite{Betal2009}. We have
observed them jointly with \nustar\ and \xmm\ during the summer of
2014, always with periods of simultaneity; the observations are
summarized in Table\,\ref{obstab}.  In this Letter, we present
only the \nustar\ and the \xmm\ EPIC data.

\nustar\ consists of two co-aligned telescopes and the two focal plane
modules, FPMA and FPMB. We reduced the \nustar\ data using the \nustar\ Data
Analysis Software (NSuTARDAS v1.4.1) available as a part of HEASOFT v6.16
and latest CALDB files. The \nustar\ observations were taken during
intervals of normal Solar activity, and we used standard filtering to
remove periods of high background during South Atlantic Anomaly (SAA)
passages and Earth occultation.
%We created cleaned,
%calibrated event files using the {\tt nupipeline} script with standard
%settings.
For each source, we used a 70\arcsec\ radius circular region around the
known position, and a 100\arcsec\ radius, circular, source-free background
region on the same detector. We generated the spectra, corresponding
response files and light curves using {\tt nuproducts}.  We grouped all
spectra using {\tt grppha} to have at least 25 counts per bin, and used
the $\chi^2$ statistic in spectral fitting. We produced background-subtracted,
barycenter-corrected light curves in the 3--10 keV and 10--30 keV
energy bands for all three IPs.

We processed the \xmm\ data with \textsc{SAS} version 13.0.0 using the latest
calibration files available in November 2014. All observations were performed
with the EPIC cameras (PN and two MOS) in small window imaging mode to avoid
pile-up, with the thin filter applied (with the exception of V709~Cas, where
the medium filter was used).  We confirmed that none of the observations was
affected by pile-up following the thread procedure\footnote{
http://xmm.esac.esa.int/sas/current/documentation/threads/epatplot.shtml}.
Then, we extracted the source spectrum from a circular region of radius
40\arcsec, centered at the known source position. The background was
obtained from a nearby source-free region of the same size for PN and
150\arcsec\ for the MOSs. For the spectral analysis, we cleaned the
observation of V709~Cas that was affected by background flares, by
applying an intensity threshold, thereby retaining only the first
$\sim22$ ks. We grouped spectral channels with the {\tt specgroup}
tool so as to have a minimum of 25 and 15 counts per bin for the PN and MOSs,
respectively.  For the timing analysis, we applied the barycentric correction
using {\tt barycen} and produced background subtracted light curves in the
0.3--3, and 3--10 keV ranges using the task {\tt epiclccorr}.

\section{Results}

\subsection{Phenomenological Fits}

Initially, we concentrated on what we could infer purely from the
\nustar\ data.  For this, we excluded the Fe K region (5.5--8.0 keV),
and the data below 4.5 keV, and approximated the interstellar plus intrinsic
absorber using a single {\tt phabs} model. To fit the continuum, we
used 3 different models with different degrees of curvature: a power law,
a cooling flow model ({\tt mkcflow}), and a single-temperature Bremsstrahlung.
We then modeled reflection using the convolution model {\tt reflect}
\citep{MZ1995}, with the abundances fixed to solar as determined by
\cite{AG1989}. We also fixed the inclination angle of the reflecting
surface at the default value of cos $\mu=0.45$ ($\mu \sim 63.3^\circ$)
(see \S 3.2). We fitted the data once with the reflection amplitude fixed
to 0.0, and once with the amplitude thawed.

\begin{figure}
%\epsscale{.80}
\plotone{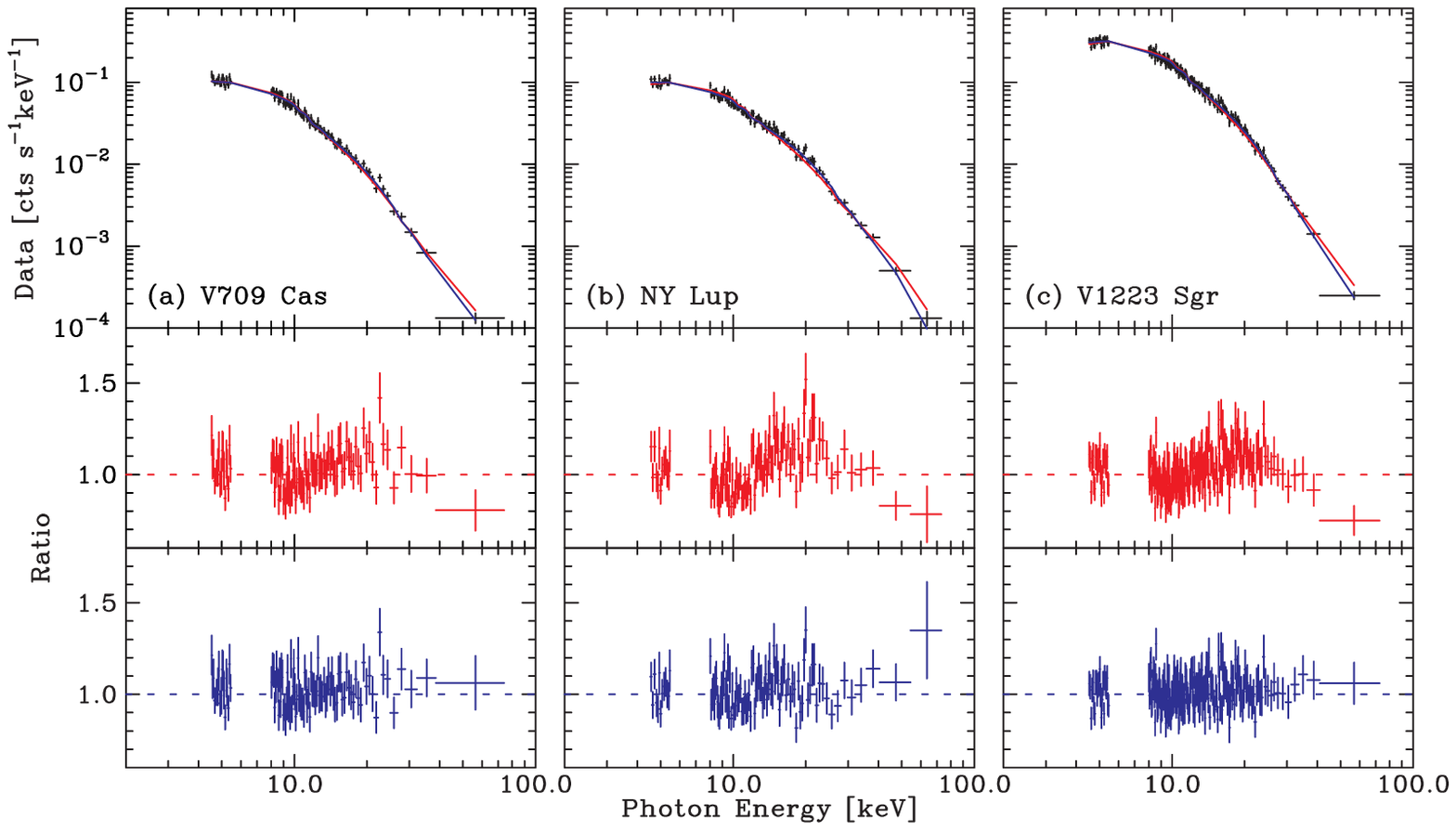}
\caption{Bremsstrahlung fits to \nustar\ spectra of 3 IPs with and without
reflection. In the top panels, data are shown along with best-fit models
with (blue) and without (red) reflection. The two lower panels show
data to model ratios for fits without (middle) and with (bottom)
reflection.\label{nuonly}}
\end{figure}

In Table\,\ref{restab}, we report the best-fit $\chi^2_\nu$ and the
degree-of-freedom of the latter case, the difference in $\chi^2$
($\Delta\chi^2$) compared to the no reflection case, and the best fit
reflection amplitude with the 90\% confidence error range.  The choice
of continuum model affects the $\Delta\chi^2$ values, with the 
Bremsstrahlung continuum presents the stringent test of reflection
(the smallest $\Delta\chi^2$). Nevertheless, reflection is detected
with high significance for all 3 models and for all 3 objects (at
a significance of 1$-6\times 10^{-10}$, for V709~Cas fitted with
Bremsstrahlung model, or higher, using the F-test).  We also
illustrate the improvement visually in (Figure\,\ref{nuonly}) using
the single-temperature Bremsstrahlung case.  The residuals against
the reflection-less model (middle panels) clearly show the continuum
reflection hump, peaking around 20 keV. We are not aware of any explanation
of the high energy end ($\sim$30 keV) of the observed hump other than
Compton reflection.

We conclude that, with \nustar\ data alone, we have achieved an
unambiguous detection of reflection in all three IPs.  However,
intrinsic absorbers in IPs are complex and often include components
with \nh\ $\sim$a few times 10$^{23}$ cm$^{-2}$. Such absorbers affect
the observed continuum up to $\sim$6 keV and also create an Fe K edge
at 7.0 keV \citep{EI1999}. We must determine the depth of the Fe K edge
due to these Compton-thin absorbers before we can place quantitative
constraints on the reflection amplitude (see \S 3.2).

We also performed simple phenomenological fits to the \xmm\ EPIC
spectra in the 5--9 keV region, where multiple emission lines are evident.
Using a Bremsstrahlung continuum modified by a single {\tt phabs} absorber,
we clearly detected the H-like (7.0 keV), He-like (6.7 keV), and near
neutral, fluorescent (6.4 keV) lines in all 3 objects, which is typical
for IPs \citep{HM2004}.  The first two indicate the presence of kT$\sim$10
keV and $\sim$5 keV plasmas, respectively. The equivalent width of the 6.4
keV line was found to be 105$\pm$11 eV, 132$\pm$12 eV and 90 $\pm$18 eV,
respectively, for V709~Cas, NY~Lup, and V1223~Sgr.

\subsection{Fits to Combined Data}

Any absorbers capable of producing an Fe K edge will necessarily
absorb soft X-rays below 2 keV almost completely. On the other hand,
many IPs often exhibit additional spectral complexities below 1 keV,
such as the soft, blackbody-like component (see, e.g., \citealt{Betal2012})
and the complex absorber may be ionized (see, e.g., \citealt{dMetal2008}).
In this initial study, we have therefore opted to fit \xmm\ EPIC (pn, MOS1
and MOS2) spectra in the 1.5--10 keV range simultaneously with the
\nustar\ (both modules) spectra in the 3.5--78 keV range. Our preliminary
analysis down to 0.3 keV confirms that the choice of low energy threshold
does not significantly affect the depth of the Fe K edge in the model.

For these joint fits, we used two different combinations of models.
In the first case (``2T''), we approximated the underlying multi-temperature
thermal emission, evidenced by the presence of H-like and He-like Fe lines,
using two {\tt mekal} \citep{Metal1985,Letal1995} components of different
temperatures. A single absorber is clearly inadequate to describe the data,
so we approximated the complex absorber using one fully covering and one
partial-covering absorbers. We applied the {\tt reflect} model to the 2
{\tt mekal} models and added a Gaussian to represent the 6.4 keV line
of cold iron\footnote{In {\tt xspec} notation, this model can be written as \\
{\tt constant*phabs(partcov*phabs)(reflect(mekal+mekal)+gaussian)}
where the constant allows for cross-normalization constant, and
the convolution model {\tt partcov} is used to create a partial-covering
absorber model based on {\tt phabs}.}.

In the second case (``mkcflow''), we modeled the underlying multi-temperature
plasma emission using the isobaric cooling flow model, {\tt mkcflow}
\citep{MS1988,Metal2003}. Although we expect post-shock plasma to cool
to the photospheric temperature of the primary, the data above 1.5 keV
have no sensitivity to plasma temperatures well below 1 keV. We therefore used
a fixed minimum temperature of 0.0808 keV, the lowest value allowed by
the {\tt mkcflow} model. The intrinsic absorption in IPs is likely to
occur in the immediate pre-shock flow, which is poorly approximated by
one or several discrete partial-covering absorbers: \cite{DM1998} pointed
this out and provided a complex absorber model with an power-law distribution
of covering fraction as a function of \nh, {\tt pwab}. We used
{\tt pwab} with minimum \nh\ set to 1.0$\times 10^{15}$ cm$^{-2}$
(effectively 0.0) as well as a {\tt phabs} to model the interstellar
absorption\footnote{This model
can be expressed as {\tt constant*phabs*pwab(reflect*mkcflow+gaussian)}
in {\tt xspec} notation.}.

In both these models, we allowed the overall abundances of the thermal
emission to vary (linked between the two {\tt mekal} components
for the ``2T'' model), which is necessary to reproduce the strengths of
the H-like and He-like Fe lines.  Moreover, we assumed that the reflection
was from the part of the white dwarf surface that is covered by freshly
accreted material, with the same abundances as the X-ray emitting plasma.
We initially fixed the angle of reflection at cos $\mu$=0.45 as before.

\begin{figure}
\plotone{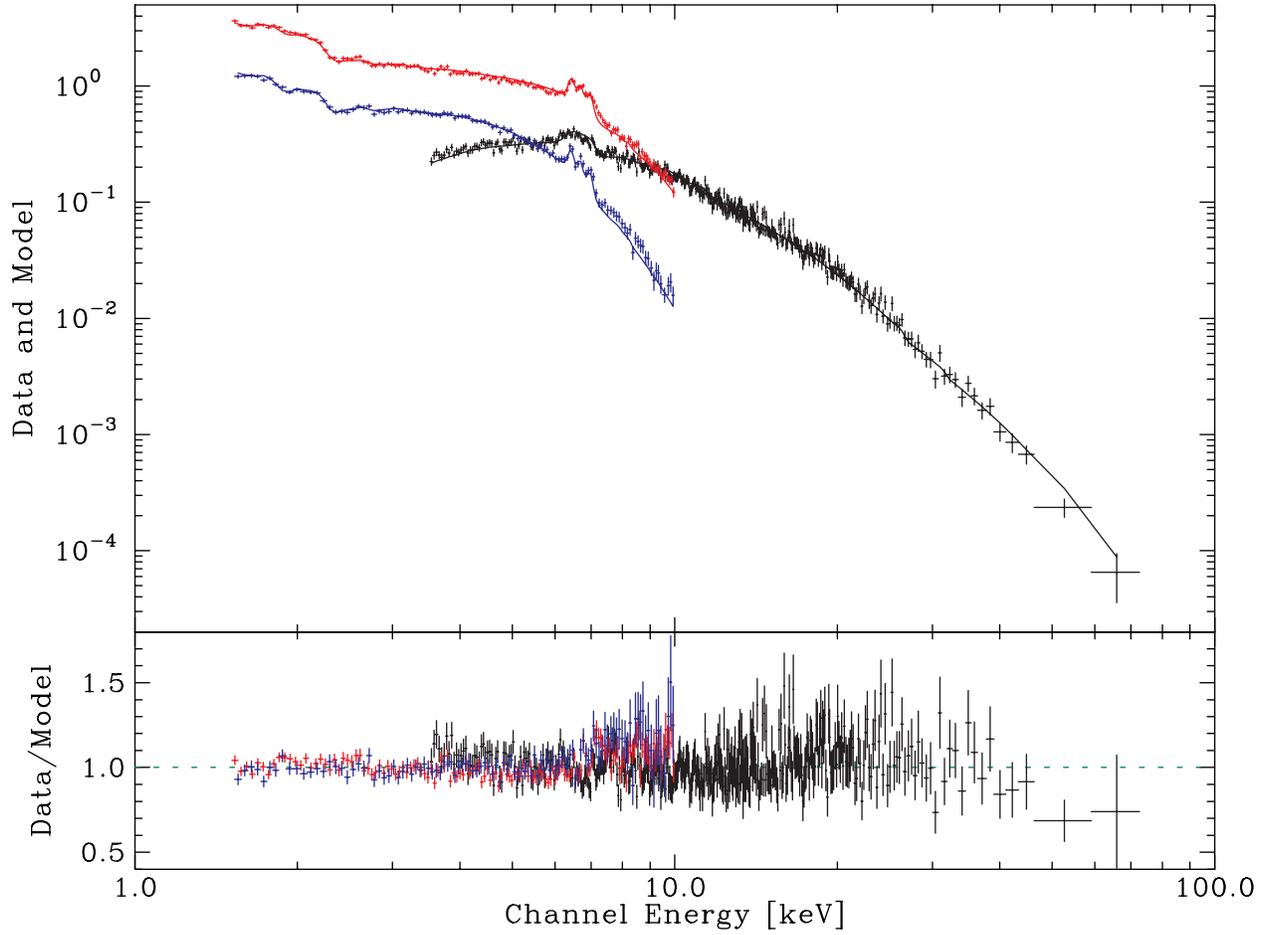}
\caption{Cooling flow fit to the combined \nustar\ and \xmm\ spectra of
V1223 Sgr. Only \nustar\ FPMA (black), \xmm\ EPIC PN (red) and MOS1 (blue)
data are shown for clarity.\label{nux}}
\end{figure}

We allowed the cross-normalization factor relative to \nustar\ FPMA
to vary, to allow for source variability and for the current limitation
in the cross-calibration of these instruments \citep{IACHEC2014,Metal2015}.
The best-fit values were all found to be within a few percent of 1.0.
However, we found several specific manifestations of cross-calibration
limitations at, e.g., 7--10 keV. This is the likely reason why the
joint \nustar/\xmm\ fits results in with $\chi^2_\nu>1.2$, which might
otherwise suggest that these fits are unacceptable. We summarize the
reflection amplitudes and other selected results in Table\,\ref{restab},
and show the results for V1223~Sgr in Figure\,\ref{nux}.

The derived reflection amplitudes are lower than in \nustar-only fits,
because our full fits indicate the presence of a complex intrinsic
absorber up to \nh\ $\sim$a few times 10$^{23}$ cm$^{-2}$, which
the \nustar-only fits did not account for. Correspondingly, these fits
result in smaller but still statistically significant values of
$\Delta\chi^2$. We also note that there is good agreement in the
reflection amplitude between the two models for V709~Cas and for
V1223~Sgr, and that they are significantly less than 1.0: $\sim$0.65
(0.45--0.9) for V1223~Sgr and $\sim$0.35 (0.1--0.6) for V709~Cas.
For NY~Lup, the different models disagree but the reflection amplitude
appears to be $>$0.8, possibly as high as 1.8. Alternatively, if we allow
for different viewing geometry, we can obtain fits with similar quality
with reflection amplitudes $\sim$1. For NY~Lup fitted with {\tt mkcflow},
this can be achieved with $\mu \sim 30^\circ$, resulting in a reflection
amplitude of 0.83--1.20. For V709~Cas and V1233~Sgr, the data can be
reconciled with reflection amplitudes of $\sim$1.0 only with edge-on
($\mu > 70^\circ$) geometries. Finally, we note that we do not confirm
the previous claim of an ionized Fe K edge at $\sim$8.0 keV in V709~Cas
\citep{dMetal2001}.

\subsection{High Energy Spin Modulation}

\begin{figure}
\plotone{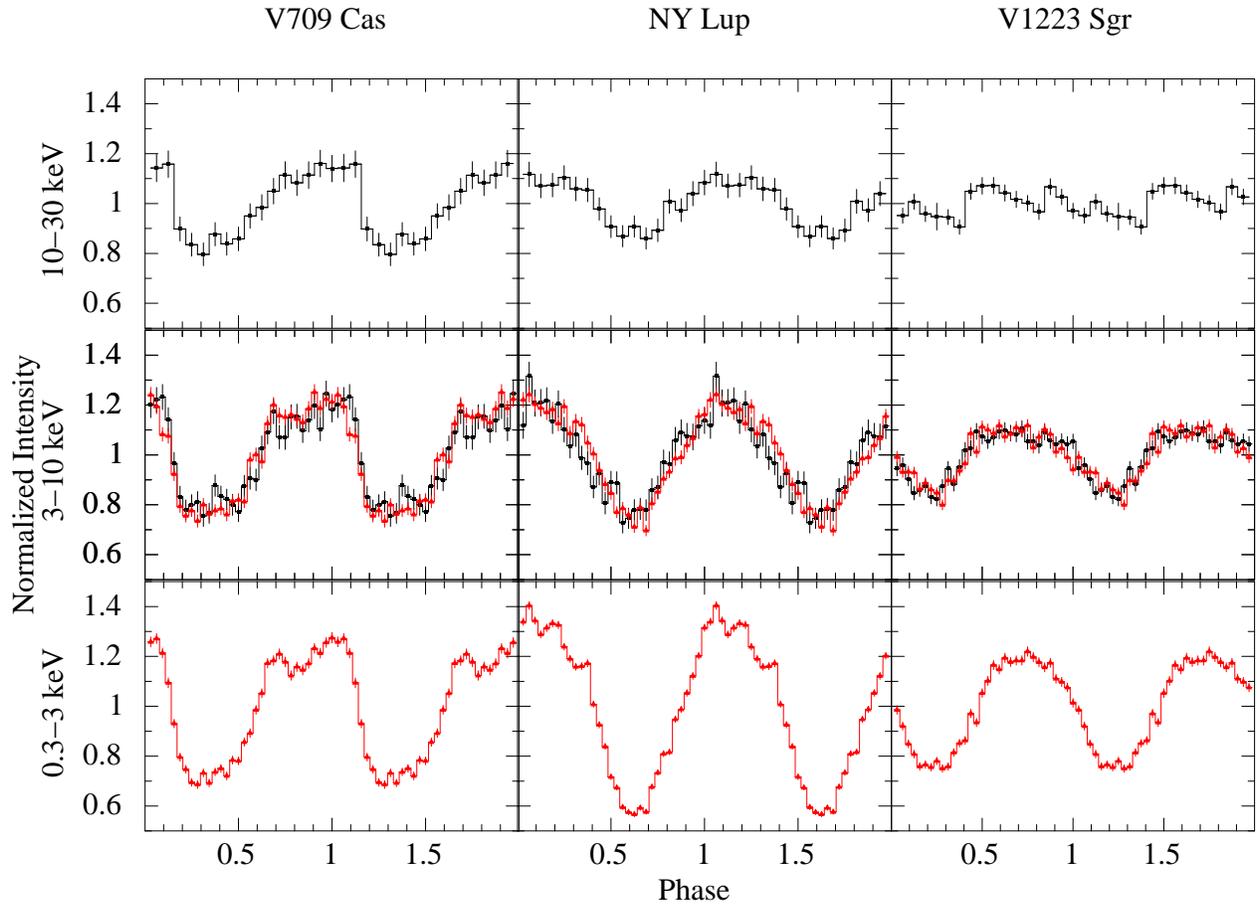}
\caption{Spin-folded light curves in 3 energy bands. Two cycles are shown.
Data used are obtained with \xmm\ EPIC PN (shown in red: 0.3--3 keV and
3--10 keV) and \nustar\ FPMA (black: 3--10 keV and 10--30 keV).\label{spin}}
\end{figure}

Energy-dependent X-ray spin modulation is a defining characteristic
of IPs, and is generally understood as due to variable complex absorber
\citep{NW1989}. This mechanism predicts small or negligible spin amplitudes
above 10 keV.  The only previous claim of spin modulation above 10 keV
is for V709~Cas, which shows
$>$10 keV modulation in phase with the lower-energy modulation in the
non-imaging \xte\ and \sax\ (PDS) data \citep{dMetal2001}.
To confirm this earlier results, and to search for subtle spin
modulation in the other objects, we folded the \xmm\ and \nustar\
light curves in three energy bands (Figure\,\ref{spin}) on the known
spin periods of 312.75 s for V709~Cas \citep{dMetal2001}, 693.01 s
for NY~Lup \citep{dMetal2006}, and 745.63 s for V1223~Sgr \citep{Oetal1985},
using the integer part of Modified Julian Date of the starting time
of each observation as the epoch of phase 0.0.

We found the pulsed fractions of 10--30 keV spin modulation to be
17.0$\pm$1.7\%, 11.6$\pm$1.6\%, and 4.2$\pm$1.1\%, respectively,
for V709~Cas, NY~Lup, and V1223~Sgr, using the definition in, e.g.,
\cite{FBetal2009}.  Our result for V709~Cas is consistent with that
of \cite{dMetal2001}. In NY~Lup, the 10--30 keV spin modulation
is roughly sinusoidal and in phase with the modulation below 10 keV;
The 10--30 keV modulation of V1223~Sgr is the weakest and non-sinusoidal
(Figure\,\ref{spin}).

\section{Discussion}

% We find that the best-fit temperatures are unaffected by the inclusion
% of reflection in some cases, but change greatly in others
% (Table\,\ref{restab}). Therefore, the reliability of white dwarf mass
% determination from the shock temperature derived without considering
% reflection (e.g., \citealt{Yetal2010}) should be assessed on a case
% by case basis.

The reflection signal is expected to be the strongest normal to the
reflecting surface, even before taking projection effect into account
\citep{GF1991}, with limited energy dependence. \cite{MZ1995}
chose cos $\mu=0.45$ as the default for their {\tt xspec} model as having
a spectral shape close to the ensemble average.  In our fiducial fits, we
used this default and found the reflection amplitude to be $<$1.0 in V709~Cas
and V1223~Sgr, and high (quite possibly $>$1.0) in NY~Lup. However, the primary
surface can cover at most 2$\pi$ steradian of sky. Our data on NY~Lup
can be reconciled with this if the viewing geometry is more face-on
($\mu \sim 30^\circ$). This would be the case if the binary inclination
($i$) is $\sim 30^\circ$ and the co-latitude of the emission
region ($\beta$) is small, which we believe is plausible for NY~Lup.

If we force the reflection amplitude to be 1.0 in V709~Cas and V1223~Sgr,
the spin-averaged $\mu$ must be $>70^\circ$. This would require either
a nearly aligned rotator seen edge-on or equatorial emission regions
seen pole-on. However, our data are also consistent with non-negligible
shock height. Unfortunately, given the abundance and \nh\ values we infer,
we cannot make a strong inference on the importance of reflection based
on the equivalent widths of the 6.4 keV lines (see Figure 7 of
\citealt{EI1999}). Therefore, we use the spin modulations to break
this degeneracy.

The soft X-rays ($<$10 keV) are modulated predominantly due to
photoelectric absorption and are at a minimum when the upper pole
(the pole on the hemisphere nearer the Earth) points most directly
towards us \citep{Hetal1987}.  In contrast, the $\mu$ dependence
of reflection should produce a hard X-ray maximum at this phase,
resulting in anti-phased spin modulations between the soft and
the hard bands. Since this is contrary to what we observe (\S 3.3),
the hard X-ray spin modulations have a different mechanism.

For NY~Lup and V1223~Sgr, the observed hard X-ray spin amplitudes
are of order 10\% or less. The complex absorbers we infer
for these objects (Table\,\ref{restab}) may produce a spin modulation
of such amplitudes via Compton scattering (Figure 6 of \citealt{R1992}),
provided that the angle-dependent reflection effects are small
(i.e., the range of $\mu$ must be small).

Finite shock height is another potential mechanism for spin modulation
\citep{M1999}. If the shock height is negligible, as is often
assumed in studies of reflection and spin modulations, the horizon
is 90$^\circ$ from zenith and the primary surface covers 2$\pi$
steradian, implying a reflection amplitude of 1.0. Moreover, in a pure
dipole geometry, one and only one pole is observed at all phases,
as the disappearance of one behind the primary rim exactly coincides
with the the appearance of the other. Once we relax this assumption,
the horizon is $>$90$^\circ$ from zenith, the surface covers
$< 2\pi$ steradian, and there is a range of viewing angles
where the X-rays from both emission regions are simultaneously observable.
This leads to an energy-independent spin modulation, since this effect
is purely geometrical.

To explain the hard X-ray spin modulation of V709~Cas, already known
to be non-sinusoidal from \rosat\ data \citep{Netal1999},
\cite{dMetal2001} proposed that its shock height is $\sim$0.2 white dwarf
radii (R$_{wd}$; see their Figure 9). Such an X-ray source has a horizon
angle of 123.5$^\circ$, both poles are observable at the viewing angles
of 66.5$^\circ$--123.5$^\circ$, and the reflection amplitude is 0.45.
This scenario provides an explanation for our observation that the
reflection amplitude is low, the 6.4 keV line is relatively weak, and
the spin modulation is even above 10 keV.

The shock in NY~Lup appears much closer to the primary surface.
As for V1223~Sgr, a characteristic height of 0.05 R$_{wd}$ would
result in a reflection amplitude of 0.7, consistent with the data
assuming a non-edge-on viewing geometry, and a horizon angle of 107.7$^\circ$.
The geometry of V1223~Sgr is poorly known; we believe that there are
plausible combinations of $i$ and $\beta$ (e.g., $\sim 50^\circ$ and
$\sim 20^\circ$, respectively) such that the lower pole is never visible.
We can then explian the observed spin modulation as due primarily t
the Compton scattering mechanism.

\section{Conclusions}

We have unambiguously detected the continuum reflection bump in
all three IPs we observed jointly with \nustar\ and \xmm.
For two of them, V709~Cas and V1223~Sgr, either we view the reflecting
surface edge-on or the reflection amplitude is low. Considering also the
high energy spin modulation we detect in V709~Cas, we argue that a post-shock
region with a significant height above the white dwarf surface can explain
all available data. We plan to continue our analysis using the \xmm\ RGS
and OM data, using EPIC data down to the lowest energy, and performing
spin-phase resolved spectroscopy of EPIC and \nustar\ data, to investigate
if this framework can provide a quantitative description of the data.

\section*{Acknowledgments}

This research has made use of data obtained with the\nustar\ mission,
a project led by the  California Institute of Technology (Caltech),
managed by the Jet Propulsion Laboratory (JPL) and funded by NASA.
We acknowledge financial support from NASA under\xmm\ grant NNX15AK63G and
from ASI/INAF contract I/037/12/0.

\clearpage

%% the file, uncomment the \plotone, \plottwo, and
%% If you need a layout that cannot be achieved with \plotone or
%% \plottwo, you can invoke the graphicx package directly with the
%% \includegraphics command or use \plotfiddle. For more information,
%% please see the tutorial on "Using Electronic Art with AASTeX" in the
%% documentation section at the AASTeX Web site, http://aastex.aas.org/
%%
%% The examples below also include sample markup for submission of
%% supplemental electronic materials. As always, be sure to check
%% the instructions to authors for the journal you are submitting to
%% for specific submissions guidelines as they vary from

%% This example uses \plotone to include an EPS file scaled to
%% 80% of its natural size with \epsscale. Its caption
%% has been written to indicate that additional figure parts will be
%% available in the electronic journal.

%\clearpage

%% Here we use \plottwo to present two versions of the same figure,
%% one in black and white for print the other in RGB color
%% for online presentation. Note that the caption indicates
%% that a color version of the figure will be available online.
%%

\clearpage

%% Tables should be submitted one per page, so put a \clearpage before

%% Two options are available to the author for producing tables:  the
%% deluxetable environment provided by the AASTeX package or the LaTeX
%% table environment.  Use of deluxetable is preferred.
%%

%% Three table samples follow, two marked up in the deluxetable environment,
%% one marked up as a LaTeX table.

%% In this first example, note that the \tabletypesize{}
%% command has been used to reduce the font size of the table.
%% We also use the \rotate command to rotate the table to
%% landscape orientation since it is very wide even at the
%% Note also that the \label command needs to be placed

%% This table also includes a table comment indicating that the full
%% version will be available in machine-readable format in the electronic

\clearpage

\begin{deluxetable}{lrcccccc}
\tabletypesize{\footnotesize}
\tablecaption{Observservations\label{obstab}}
\tablehead{
\colhead{Object} & \colhead{Obs. Date}
 & \multicolumn{2}{c}{\nustar} & \multicolumn{2}{c}{pn}
 & \multicolumn{2}{c}{MOS} \\
 & & \colhead{Start} & \colhead{Exp.}
 & \colhead{Start} & \colhead{Exp.} & \colhead{Start} & \colhead{Exp.} \\
 & & \colhead{Time} & \colhead{(ks)}
 & \colhead{Time} & \colhead{(ks)} & \colhead{Time} & \colhead{(ks)} \\
}
\startdata
V709 Cas  & 2014-07-07 & 02:01 & 26 & 04:43 & 23 & 04:37 & 31 \\
NY Lup    & 2014-08-09 & 14:51 & 23 & 08:19 & 26 & 08:13 & 36 \\
V1223 Sgr & 2014-09-16 & 02:26 & 20 & 02:17 & 12 & 02:11 & 16 \\
\enddata
\end{deluxetable}

\clearpage

\begin{deluxetable}{lrrrrrrcccc}
\tabletypesize{\footnotesize}
\rotate
\tablecaption{Selected Spectral Fit Results\label{restab}}
\tablehead{
\colhead{Object} & \colhead{Data} & \colhead{Model\tablenotemark{a}} &
\colhead{$\chi^2_\nu$/DOF} & \colhead{$\Delta\chi^2$\tablenotemark{b}} & \colhead{Amplitude} &
\colhead{N$_{\rm H}$\tablenotemark{c}} & \colhead{PCF/$\alpha$\tablenotemark{c}} &
\colhead{kT\tablenotemark{d}} & \colhead{Abund\tablenotemark{e}} \\
 & & & & & & \colhead{10$^{23}$\,cm$^{-2}$} & & \colhead{keV} & \\
}
\startdata
V709 Cas  & \nustar & PL      & 1.16/557  & 78.9  & 2.16 (1.57--2.91) & & & & \\
          &         & mkcflow & 1.02/557  & 81.7  & 1.71 (1.27--2.24) & & & & \\
          &         & Brems   & 1.05/557  & 41.7  & 0.80 (0.56--1.07) & & & & \\
          & Joint   & 2T      & 1.29/1094 &  8.7  & 0.27 (0.11--0.44) & 1.5 & 31\% & 27.0 (28.6) & 0.34 \\
          &         & mkcflow & 1.30/1096 & 12.4  & 0.38 (0.21--0.60) & 4.7 & $-$0.84 & 46.6 (48.2) & 0.28 \\

NY Lup    & \nustar & PL      & 1.02/401  & 219.4 & 3.95 (3.07--5.12) & & & & \\
          &         & mkcflow & 0.97/401  & 279.4 & 2.68 (2.13--3.26) & & & & \\
          &         & Brems   & 1.00/401  & 146.0 & 1.47 (1.23--1.75) & & & & \\
          & Joint   & 2T      & 1.20/937  &  99.1 & 1.02 (0.79--1.15) & 1.6 & 49\% & 38.9 (56.6) & 0.76 \\
          &         & mkcflow & 1.30/939  &  40.0 & 1.62 (1.53--1.83) & 4.7 & $-$0.78 & 54.3 (37.6) & 0.68 \\

V1223 Sgr & \nustar & PL      & 1.30/599  & 233.5 & 2.81 (2.30--3.41) & & & & \\
          &         & mkcflow & 0.95/599  & 222.5 & 2.00 (1.67--2.39) & & & & \\
          &         & Brems   & 0.97/599  & 119.5 & 1.13 (0.91--1.38) & & & & \\
          & Joint   & 2T      & 1.37/1141 &  71.1 & 0.62 (0.46--0.78) & 1.4 & 48\% & 22.2 (23.5) & 0.33 \\
          &         & mkcflow & 1.37/1143 &  43.6 & 0.66 (0.46--0.79) & 4.5 & $-$0.69 & 34.3 (34.2) & 0.26 \\

\enddata
\tablenotetext{a}{See text for full descriptions of the models.}
\tablenotetext{b}{Difference in $\chi^2$ between fits with reflection amplitude
      let free to fit and fixed to 0.0}
\tablenotetext{c}{The column density for the partial covering absorber and
      the partial covering fraction (``2T''), or the maximum
      column and the power-law index for of the {\tt pwab} absorber
      (``mkcflow''). }
\tablenotetext{d}{Temperature of the high temperature component (``2T'') or
      the highest temperature (``mkcflow'') in the best-fit
      model, with the best-fit value with reflection amplitude set to 0.0
      in parentheses. }
\tablenotetext{e}{Overall abundance of accreting plasma, assumed to be
      responsible for both emission and reflection.}
\end{deluxetable}

\end{document}